\documentclass{elsart}

\usepackage{natbib}

 \usepackage{graphicx}

\usepackage{amssymb}

\begin{document}

\begin{frontmatter}

\title{Three-dimensional MHD simulations of X-ray emitting subcluster plasmas in cluster of galaxies}

\author[na]{N. Asai\corauthref{cor}},
\corauth[cor]{Corresponding author.}
\ead{asai@astro.s.chiba-u.ac.jp}
\author[nf]{N. Fukuda},
\ead{fukudany@sp.ous.ac.jp}
\author[rm]{R. Matsumoto}
\ead{matumoto@astro.s.chiba-u.ac.jp}

\address[na]{Graduate School of Science and Technology, Chiba University, 
1-33 Yayoi-cho, Inage-ku, Chiba 263-8522, Japan}

\address[nf]{Department of Computer Simulation, Faculty of Informatics, 
Okayama University of Science, 1-1 Ridai-cho, Okayama 700-0005, Japan}

\address[rm]{Department of Physics, Faculty of Science, Chiba University, 
1-33 Yayoi-cho, Inage-ku, Chiba 263-8522, Japan}

\begin{abstract}

Recent high resolution observations by the {\it Chandra} X-ray satellite revealed various
substructures in hot X-ray emitting plasmas in cluster of galaxies. For example,
{\it Chandra} revealed the existence of sharp discontinuities in the surface brightness at the
leading edge of subclusters in merging clusters (e.g., Abell 3667), where the temperature
drops sharply across the fronts. These sharp edges are called cold fronts.
We present results of three-dimensional (3D) magnetohydrodynamic simulations of the interaction 
between a dense subcluster plasma and ambient magnetized intracluster 
medium. Anisotropic heat conduction along magnetic field lines is included. 
At the initial state, magnetic fields are assumed to be uniform and transverse to 
the motion of the dense subcluster.
Since magnetic fields ahead of the subcluster slip toward the third direction 
in the 3D case, the strength of magnetic fields in this region can be reduced compared to that in the 
2D case.
Nevertheless, a cold front can be maintained because the magnetic field lines wrapping 
around the forehead of the subcluster suppress the heat conduction across them. 
On the other hand, when the magnetic field is absent, a cold front cannot be maintained 
because isotropic heat conduction from the hot ambient plasma rapidly heats 
the cold subcluster plasma.

\end{abstract}


\end{frontmatter}

\section{Introduction}\label{sec:intro}

In recent years, high spatial resolution observations of clusters of galaxies by 
{\it Chandra} revealed the existence of sharp discontinuities of X-ray intensity 
called cold fronts \citep[e.g.,][]{mar00,vik01b}.
Their existence is challenging because in high temperature plasma, 
high thermal conductivity quickly smears such a discontinuity in temperature.
\citet{ett00} pointed out that heat conductivity across the front should be 
reduced from the classical Spitzer value, $\kappa_{\rm Sp}= 5 \times 10^{-7}T^{5/2} \, 
{\rm erg \, s^{-1} \, cm^{-1} \, K^{-1}}$ \citep{spi62} by several orders of magnitude.
Magnetic fields enable such suppression.
Cold fronts may also be subject to the Kelvin-Helmholtz (K-H) instability.
\citet{vik01a,vik01b} suggested that turbulent magnetic fields stretched 
along the cold front will reduce the growth rate of the K-H instability 
at the interface between a subcluster plasma and an ambient plasma.

The magnetic field strength in the core of clusters of galaxies can be the order of ${\rm \mu G}$ 
\citep[e.g.,][]{kro94,car02}. \citet{joh04} summarized the observations of 
the field strength in several regions of Abell 3667 by using different
techniques. They showed that its strength in the cluster core is  about $1-2 \, {\rm \mu G}$,
and that magnetic fields are tangled on scales of roughly $100 \, {\rm kpc}$. 

Let us estimate the heat conduction time scale in cluster of galaxies.
It has been suggested that the heat conduction in cluster of galaxies can be very efficient
\citep[e.g.,][]{tak77} because the Coulomb mean free path, 
$l_{\rm c} \sim 5 \, (kT / 5 \, {\rm keV})^{2}(n / 10^{-3} \, {\rm cm^{-3}})^{-1} 
\, {\rm kpc}$, is large. Here, $n$ is the number density. 
The time required for heat to diffuse by conduction across a length  
$L$ is given by $\tau_{\rm Sp} \sim \rho L^{2}/ \kappa_{\rm Sp}
\sim 3 \times 10^{7} (kT/ 5 \, {\rm keV})^{-5/2}
(n / 10^{-3} \, {\rm cm^{-3}})(L/100 \, {\rm kpc})^{2} \, {\rm yr}$, 
where $\rho$ is the density and $\kappa_{\rm Sp}$ is the Spitzer conductivity.
Since the thickness of the cold front is much smaller than $100 \, {\rm kpc}$,
they will disappear in a time scale shorter than $10^{7} \,  {\rm yr}$.
If magnetic fields exist, however, the characteristic length of the 
heat exchange across magnetic field lines is reduced to the Larmor radius, 
$r_{\rm L} = 2500 \, (B/1 \, {\rm \mu G})^{-1} (kT/5 \, {\rm keV})^{1/2} 
\, {\rm km}$.
Thus, the heat conduction across the magnetic field is almost suppressed.

Earlier numerical studies of cold fronts \citep[e.g.,][]{hei03} did not include 
magnetic fields and heat conduction. 
\citet{asa04} first reported the results of two-dimensional (2D) 
magnetohydrodynamic (MHD) simulations of cold fronts including both magnetic fields
and anisotropic heat conduction. They showed that magnetic fields are stretched and 
elongated ahead of the front. Since these magnetic fields suppress the heat conduction 
across them, the contact discontinuity (a cold front) between the cold subcluster plasma 
and the hot ambient plasma can be maintained.
The cold subcluster is wrapped by magnetic fields and protected from being heated 
by the heat conduction.

In this paper, we focus on the three-dimensional (3D) effects.
In the 3D case, since magnetic fields compressed in front of the cold front can expand 
in the third direction or slip out the front surface, 
the strength of the magnetic fields in this region may be reduced compared to that in the 2D case.
We would like to show that even when we include these 3D effects, cold fronts can 
be maintained.

\section{Simulation model}
\label{sec:model}

We simulated the time evolution of a cluster plasma in a 
frame comoving with the subcluster. The basic equations are as follows:
\begin{equation}
\frac{\partial \rho}{\partial t} + \mbox{\boldmath $\nabla$ $\cdot$}
 (\rho \mbox{\boldmath $v$}) = 0,
\end{equation}
\begin{equation}
\rho \left[
\frac{\partial \mbox{\boldmath $v$}}{\partial t} 
+ (\mbox{\boldmath $v$ $\cdot$ $\nabla$}) 
\mbox{\boldmath $v$}
\right] =
-\mbox{\boldmath $\nabla$}{\it p} + 
\frac{(\mbox{\boldmath $\nabla$} \times 
\mbox{\boldmath $B$}) \times \mbox{\boldmath $B$}}{4 \pi} 
- \rho \mbox{\boldmath $\nabla$} \psi, 
\end{equation}
\begin{equation}
\frac{\partial \mbox{\boldmath $B$}}{\partial t} 
= \mbox{\boldmath$\nabla$} \times (\mbox{\boldmath $v$} 
\times \mbox{\boldmath $B$}),
\end{equation}
\begin{eqnarray}
\lefteqn{
\frac{\partial}{\partial t} 
\left[
\frac{1}{2} \rho v^{2} 
+ \frac{B^{2}}{8 \pi} + \frac{p}{\gamma -1}
\right]
} 
 \\ 
& &
\lefteqn{
+ \mbox{\boldmath$\nabla$$\cdot$}
\left[
\left(\frac{1}{2} \rho v^{2} + \frac{\gamma p}{\gamma -1}
\right) \mbox{\boldmath$v$} + \frac{1}{4 \pi} (-\mbox{\boldmath $v$} 
\times \mbox{\boldmath$B$}) 
\times \mbox{\boldmath $B$} - \kappa_{\parallel}  
\mbox{\boldmath $\nabla_{\parallel}$} T
\right] 
= - \rho \mbox{\boldmath $v$ $\cdot$ $\nabla$} \psi,
} \nonumber
\end{eqnarray}
where $\rho$, $\mbox{\boldmath $v$}$, $p$, $\mbox{\boldmath $B$}$, and 
$\psi$ are the density, velocity, pressure, magnetic fields, 
and gravitational potential, respectively. We use the specific heat ratio 
$\gamma = 5/3$. The subscript $\parallel$ denotes the components parallel 
to the magnetic field lines. We assume that heat is conducted only 
along the field lines. We solved ideal MHD equations in a Cartesian 
coordinate system $(x, y, z)$ by a modified Lax-Wendroff method with 
artificial viscosity. The heat conduction term in the energy equation is solved 
by the implicit red and black successive overrelaxation method 
\citep[see][for detail]{yok01}. 
The radiative cooling term is not included.
The units of length, velocity, density, pressure, temperature, and time
in our simulations are
$r_{0}=250 \, {\rm kpc}$,
$v_{0}=800 \, {\rm km \, s^{-1}}$,
$\rho_{0}=5 \times 10^{-27} \, {\rm g \, cm^{-3}}$,
$p_{0} = 3 \times 10^{-11} \, {\rm erg \, cm^{-3}}$,
$kT_{0}= 4 \, {\rm keV}$, and
$t_{{0}}=r_{0}/v_{0}= 3 \times 10^{8} \, {\rm yr}$,
respectively.

\begin{figure}[t]
\begin{center}
\includegraphics*[width=7.5cm]{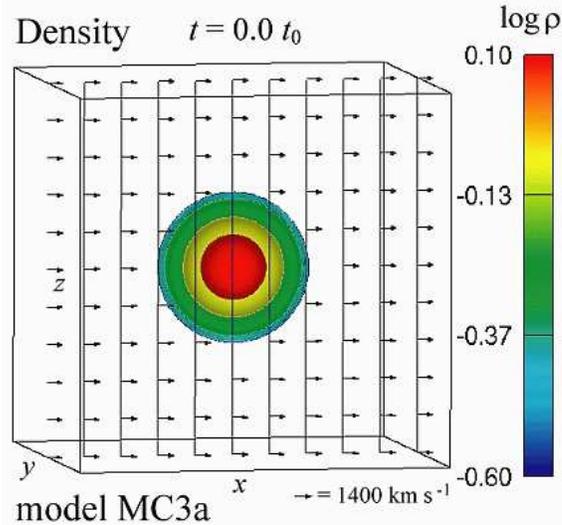}
\end{center}
\caption{Initial density distribution visualized as logarithm of the density.
}
\label{fig:f1}
\end{figure}

Figure \ref{fig:f1} shows the initial density distribution. 
Solid lines and arrows show magnetic field lines and velocity vectors in the $y=0$ plane, respectively.
We assume that a spherical isothermal low-temperature
($kT_{\rm in} = 4 \, {\rm keV}$) plasma is confined by the gravitational
potential of a subcluster. The low-temperature plasma is embedded in the 
less-dense $(\rho_{\rm out} = 0.25 \, \rho_{0})$ uniform, hot 
$(kT_{\rm out} = 8 \, {\rm keV})$ plasma. Here the subscripts ``in'' 
and ``out'' denote the values inside and outside the subcluster, respectively.

We assume that the density distribution of the subcluster is given 
by the $\beta$-model profile \citep{cav76},
\begin{equation}
\rho_{\rm in} = \rho_{\rm c}\left[1+\left(\frac{r}{r_{\rm c}} \right)^{2} \right]^{-3\beta/2}
\end{equation}
where $r = (x^{2}+y^{2}+z^{2})^{1/2}$, $\beta=2/3$, the maximum density 
$\rho_{\rm c}= 10^{-26} \, {\rm g \, cm^{-3}}$, and 
the core radius is $r_{\rm c} = 290 \, {\rm kpc}$. 
The subcluster is initially in hydrostatic equilibrium under the gravitational 
potential fixed throughout the simulation.

We assume that ambient plasma initially has a uniform speed with 
Mach number $M = v_{x}/c_{\rm s \, out} = 1$, 
where $c_{\rm s \, out}$ is the ambient sound speed. The Mach number 
with respect to the sound velocity inside the subcluster is 
$ M^{\prime}= v_{x}/c_{\rm s \, in}= \sqrt{2}$. 

Table \ref{tbl1} shows model parameters. 
An important parameter is the plasma beta ($\beta_{0}$) defined as the ratio of 
the ambient gas pressure to the magnetic pressure.
When $\beta_{0} = p_{\rm gas} / p_{\rm mag}= 100$, the magnetic field strength is
$B \sim 1.5 \, {\rm \mu G}$.
Models HC2 (2D) and HC3 (3D) are non-magnetic models with isotropic heat
conduction, and model MC2a (2D) and MC3a (3D) are models with magnetic fields 
($\beta_{0}=100$) and anisotropic heat conduction. 
Model MC2b (2D) is a model with weak magnetic fields ($\beta_{0}=1000, B \sim 
0.5 \, {\rm \mu G}$) and anisotropic heat conduction.
Model MC3b (3D) is a model with weaker magnetic fields 
($\beta_{0}=10^{3}, B \sim 0.15 \, {\rm \mu G}$).
We note that the initial magnetic fields are parallel to the $z$-axis 
and perpendicular to the motion of the subcluster.

When magnetic fields exist, heat is conducted only in the 
direction parallel to the field lines.
Along the magnetic fields, the heat conductivity is taken to be that of 
the Spitzer conductivity,
$\kappa_{\parallel} = 5 \times 10^{-7} \, T^{5/2} \, {\rm erg \, s^{-1} 
\, cm^{-1} \, K^{-1}}$. 
Meanwhile the conductivity across the field lines is $ \kappa_{\perp} = 0$. 
The conduction time scale along the field lines is 
$\tau_{\kappa} \sim \rho L^{2}/ \kappa_{\parallel}
\sim 3 \times 10^{7} \, (kT/ 5 \, {\rm keV})^{-5/2} 
(n / 10^{-3} \, {\rm cm^{-3}})\\ (L/100 \, {\rm kpc})^{2} \, {\rm yr}$. 

For boundary conditions, the left boundary at $x= -5$ is 
taken to be a fixed boundary, and other boundaries are taken to be
free boundaries where waves can be transmitted.

\begin{table}
\caption{Models and parameters. Heat conduction is isotropic in models HC2 and HC3. 
In other models, heat conducts only along magnetic fields.}
\label{tbl1}
\begin{center}
\begin{tabular}[h]{ccccccc}
\hline
\hline
Model & $\beta_{0}$ & $\kappa$ & $kT_{\rm in}$[keV] & $kT_{\rm out}$[keV]
& simulation box & number of grid points\\
\hline
HC2 & $\infty$ & $\kappa$ & 4 & 8  & ${(2.5 {\rm Mpc})}^{2}$ & $ 1200^{2}$\\
MC2a & 100 & $\kappa_{\parallel}$ & 4 & 8  & ${(2.5 {\rm Mpc})}^{2}$ & $1200^{2}$\\
MC2b & 1000 & $\kappa_{\parallel}$ & 4 & 8  & ${(2.5 {\rm Mpc})}^{2}$ & $1200^{2}$\\
HC3 & $\infty$ & $\kappa$ & 4 & 8  & ${(2.5 {\rm Mpc})}^{3}$ & $280^{3}$\\
MC3a &100 & $\kappa_{\parallel}$& 4 & 8  & ${(2.5 {\rm Mpc})}^{3}$&$280^{3}$\\
MC3b & $10^{4}$ & $\kappa_{\parallel}$& 4 & 8  & ${(2.5 {\rm Mpc})}^{3}$&$280^{3}$\\
\hline
\end{tabular}
\end{center}
\end{table}

\section{Results}\label{sec:res}

\subsection{Configuration of magnetic fields}\label{sec:res1}
 
\begin{figure}[t]
\begin{center}
\includegraphics*[width=7.5cm]{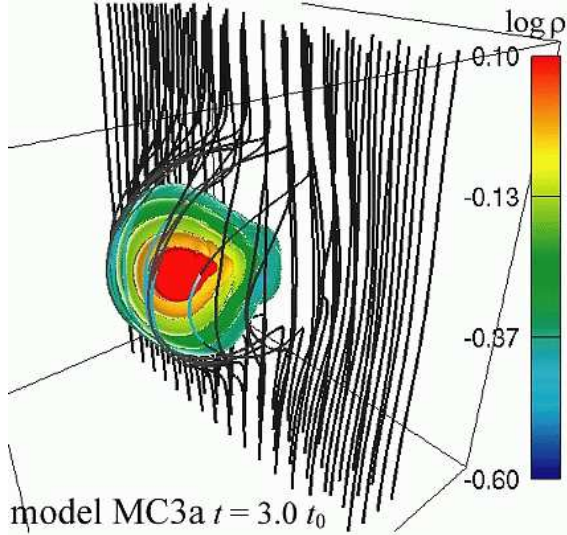}
\end{center}
\caption{Magnetic field lines at $t=1 \, {\rm Gyr}$. Color contours show the 
logarithmic density distribution.
The solid curves show magnetic field lines.}
\label{fig:f2}
\end{figure}

Figure \ref{fig:f2} shows the magnetic fields for model MC3a at $t=1 \, {\rm Gyr}$. 
Solid curves and color contours show the magnetic field lines and density 
distribution, respectively. 
The subcluster plasma is wrapped by the magnetic field lines.
Ahead of the subcluster, the magnetic field lines are almost parallel to the 
contact surface between the subcluster plasma and the ambient plasma.
We can identify magnetic field lines slipping in the $y$-direction. 

\begin{figure}[h]
\begin{center}
\includegraphics*[width=13.5cm]{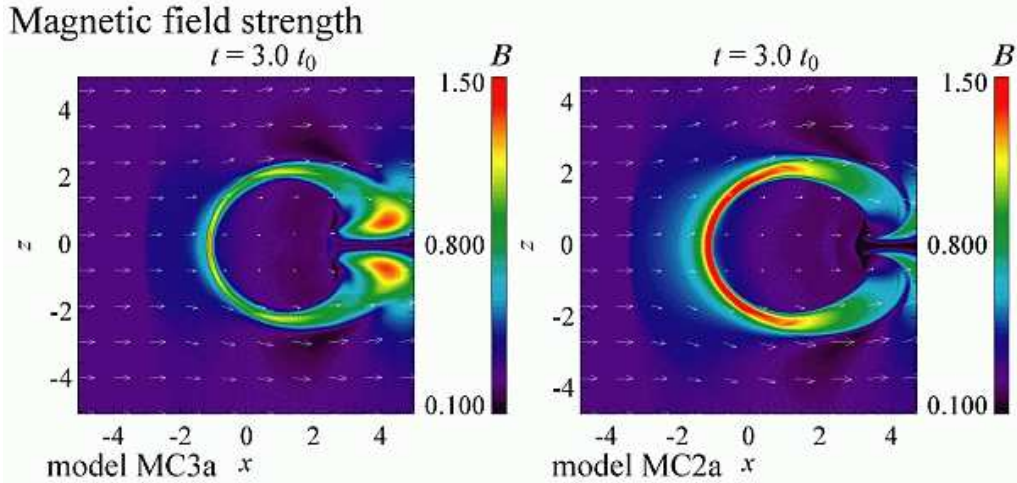}
\end{center}
\caption{Comparison of the magnetic field strength  at $t=1 \, {\rm Gyr}$ 
between 3D (left) and 2D (right). 
The left and right panels show the result for models MC3a and MC2a 
at $y=0$ plane, respectively. Arrows show the velocity vectors.
}
\label{fig:f3}
\end{figure}

Figure \ref{fig:f3} shows isocontours of magnetic field strength in the 
$y=0$ plane for models MC3a (left, 3D) and MC2a (right, 2D) at $t=1 \, {\rm Gyr}$, respectively. 
Arrows show velocity vectors.
We note that the initial magnetic field lines are parallel to the $z$-axis. 
The magnetic field strength is enhanced ahead of the subcluster via compression. 
Compared to the 2D model (right), magnetic fields in the 3D model are weaker ahead of 
the subcluster but their strength is enhanced at the tail of 
the subcluster because the plasma flows advect the magnetic fields toward the tail and 
they pile up in the tail.

\begin{figure}[h]
\begin{center}
\includegraphics*[width=13.5cm]{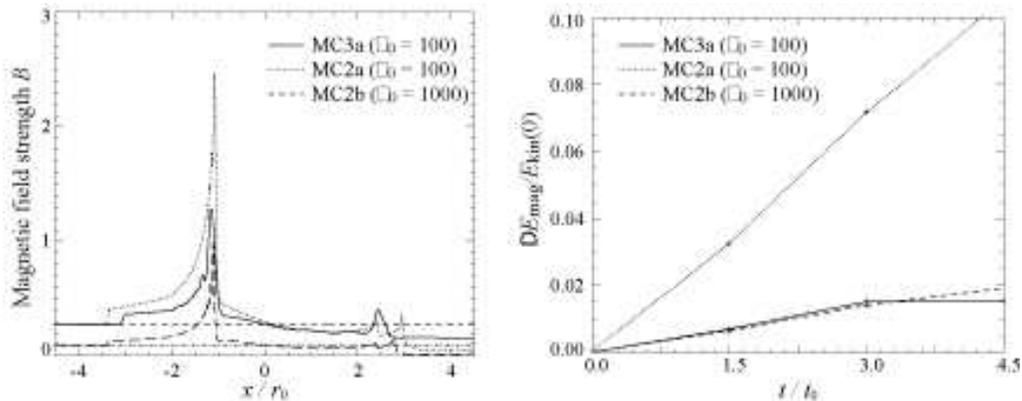}
\end{center}
\caption{
Left: Distributions of magnetic field strength at $t=1 \, {\rm Gyr}$.
The solid curve shows magnetic field strength for model MC3a (3D, $\beta_{0}=100$) 
along the $x$-axis.
The dotted curve and the long-dashed curve show the strength for model MC2a (2D, $\beta_{0}=100$)
and MC2b (2D, $\beta_{0}=1000$) along the $x$-axis, respectively.
The dashed line and dash-dotted line show the initial magnetic field strength for models 
with $\beta_{0}=100$ and $\beta_{0}=1000$, respectively.
Right: Time evolution of the magnetic energy ($E_{\rm mag}$) integrated 
in the whole simulation region. The increase with respect to the initial kinetic energy, 
$\Delta E_{\rm mag}/E_{\rm kin}(0)=(E_{\rm mag}(t)- E_{\rm mag}(0))/ E_{\rm kin}(0)$, 
is plotted. The solid line, dotted line, and long-dashed line are 
for models MC3a (3D, $\beta_{0}=100$), MC2a (2D, $\beta_{0}=100$), 
and MC2b (2D, $\beta_{0}=1000$), respectively.}
\label{fig:f4}
\end{figure}

The left panel of Figure \ref{fig:f4} shows the distribution of magnetic field 
strength at $t=1 \, {\rm Gyr}$ along the $x$-axis. 
The solid curve shows their strength for model MC3a (3D, $\beta_{0}=100$).
The dotted line and the long-dashed line show their strength for model MC2a 
(2D, $\beta_{0}=100$) and MC2b (2D, $\beta_{0}=1000$), respectively. 
Their strength in front of the cold front is enhanced more than a factor 3 from the initial state.  
The right panel in Figure \ref{fig:f4} shows the time evolution of the total magnetic 
energy integrated in the whole simulation region.
The magnetic energy increases gradually due to magnetic field compression 
around the subcluster.
 

\subsection{Effects of magnetic fields on heat conduction}\label{sec:res2}

\begin{figure}[t]
\begin{center}
\includegraphics*[width=13.5cm]{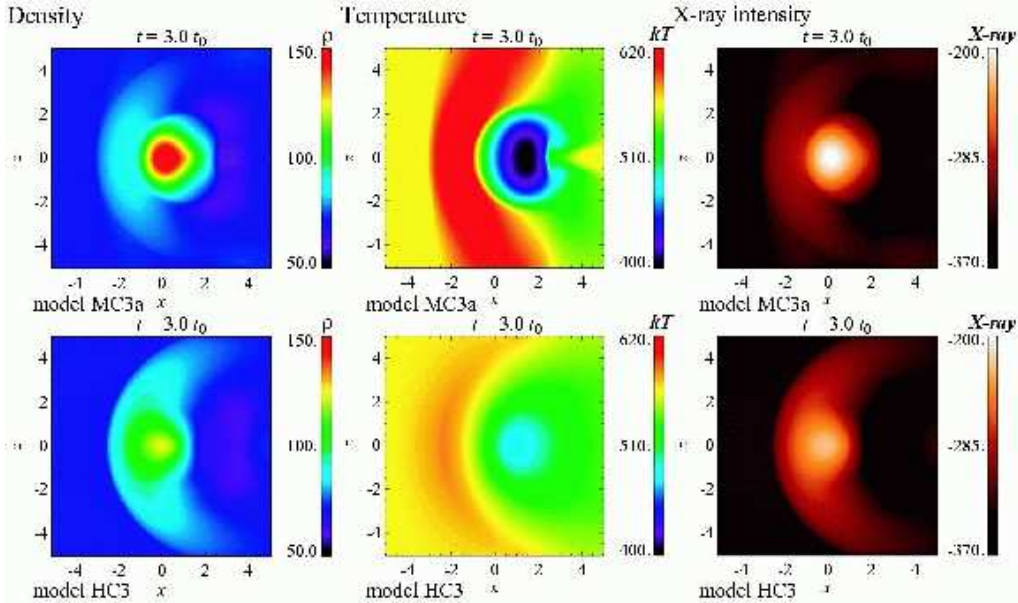}
\end{center}
\caption{Distribution of projected density (left), temperature (middle), 
and X-ray intensity (right) integrated at $t=1 {\rm Gyr}$ along the $y$-axis. 
The top panels show results for model MC3a (with magnetic fields and heat conduction). 
The bottom panels show results for model HC3 (without magnetic fields 
and with heat conduction).}
\label{fig:f5}
\end{figure}

Let us compare the results for model MC3a (with heat conduction) 
and model HC3 (without heat conduction).
We show in Figure \ref{fig:f5} distributions of the projected density, temperature, 
and X-ray intensity obtained by integrating these quantities in the $y$-direction.
X-ray intensity are visualized from simulation results as logarithm of the thermal
bremsstrahlung emissivity, $\sim \rho^{2}$.
In model HC3 (lower panels), heat is conducted isotropically. 
Since the conduction time scale across the cold front with width $\sim 5 {\rm kpc}$
\citep{vik01b} is very short ($\tau_{\kappa} \sim 10^{7} {\rm yr}$), 
the cool component inside the subcluster is heated by the hot ambient plasma and evaporates rapidly.
Meanwhile in model MC3a (top panels), heat is conducted only in the direction 
parallel to the field lines. 
As we showed in \S \ref{sec:res1}, since the subcluster is wrapped by magnetic field lines, 
the cool component inside the subcluster is protected from heating.
We can see this protection remarkably in the temperature distributions (middle panels).
Thus the discontinuity remains sharp. 

\begin{figure}[h]
\begin{center}
\includegraphics*[width=13.5cm]{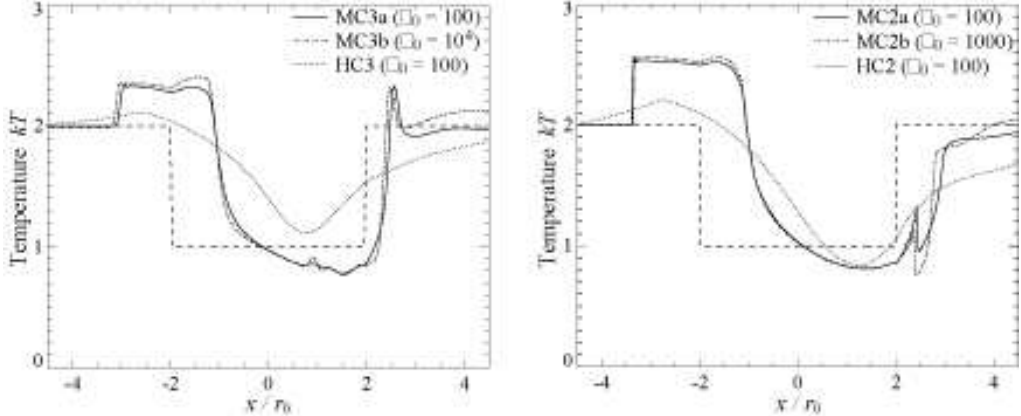}
\end{center}
\caption{Comparison of the temperature distributions between 3D (left) 
and 2D (right) at $t=1 \, {\rm Gyr}$ along the $x$-axis.
The left panel shows the results for models MC3a (solid line), MC3b (dash-dotted line),
and HC3 (dotted line).
The right panel shows the results for models MC2a (solid line), MC2b (dash-dotted line), 
and HC2 (dotted line).
The dashed lines in both panels show the initial temperature distribution.}
\label{fig:f6}
\end{figure}

Figure \ref{fig:f6} shows the temperature profiles at $t= 1 {\rm Gyr}$ along the $x$-axis.
The left panel shows those for models MC3a (solid line) and HC3 (dotted line).
The right panel shows those for models MC2a (solid line), MC2b (dash-dotted line),
and HC2 (dotted line). 
In both panels, the dashed line shows the initial temperature profile.
It is obvious that magnetic fields along the front enables the cold front to exist 
for $1 {\rm Gyr}$.

We can also identify shock waves. 
Numerical results reproduced the bow shock produced by supersonic motion 
upstream of the cold front observed in A3667 (see Figure \ref{fig:f5}). 
The interaction of the subcluster plasma with the ambient plasma generates another shock 
propageting inside the subcluster. This shock exists at $x=2.5$ in models 
MC3a and MC2a (solid lines) in Figure \ref{fig:f6}, and also visible 
in the left panel of Figure \ref{fig:f4}.
These shocks drive the adiabatic expansion of the subcluster and cools 
the subcluster plasma. Such behaviors are consistent with those in 2D models \citep{asa04}.


\section{Discussion}\label{sec:dis}

\subsection{Importance of magnetic fields for the existence of cold fronts}\label{sec:dis1}

We have shown that magnetic fields enable the existence of cold fronts by suppressing the 
heat conduction across the front. 
Previous hydrodynamic simulations \citep[e.g.,][]{hei03} could reproduce the cold fronts 
because they neglected the heat conduction.  
When heat conduction is included, cold fronts disappear 
in time scales of $\sim 10^{8} \, {\rm yr}$ unless the heat conduction is 
suppressed as we showed in models HC2 and HC3 (see Figures \ref{fig:f5} and \ref{fig:f6}).
Magnetic fields much smaller than that in model MC3a can suppress heat conduction.
We confirmed that weak magnetic fields for model MC3b ($\beta_{0}=10^{4}, 
B \sim 0.15 \, {\rm \mu G}$) can suppress the heat conduction 
across the front as long as they cover the front.
Even when large-scale coherent magnetic fields do not exist, 
turbulent magnetic fields can suppress the heat conduction across 
the front; they are stretched and elongated parallel to the front 
\citep[see also][]{vik01a,asa04}.

\citet{vik01a,vik01b} suggested that magnetic fields are important for the 
existence of cold fronts because they can suppress the growth of the K-H instability.
\citet{asa04} showed that in models without heat conduction and without magnetic fields,
the growth of the K-H instability is not prominent and does not disrupt the cold front 
ahead of the moving subcluster. In the sides of the subcluster, 
the K-H instability creates eddies which detach and propagate downstream.
Magnetic fields can suppress the formation of such eddies.

\subsection{Possibility of subcluster disruption 
by magnetic fields accumulating ahead of the subcluster}\label{sec:dis2}

Let us now discuss the possibility of the disruption of the subcluster plasma 
by magnetic fields. \citet{gre00} presented the results of 3D simulations of 
moderate supersonic cloud motion in a magnetized interstellar medium. 
They assumed that the cloud has uniform density. They showed that magnetic fields 
accumulating ahead of the cloud triggers the Rayleigh-Taylor (R-T) instability 
and disrupts the cloud. 
Our numerical results, however, does not show the growth of such R-T instability. 
These different behaviors are due to a difference of the density distribution of the cloud.
Since the subcluster plasma in our simulations has a dense core because it is confined 
by the gravity, magnetic fields ahead of the clump can not deform the density distribution 
inside the subcluster.

\subsection{Three-dimensional effects}\label{sec:dis3}

We have shown that in the 3D case, since magnetic fields ahead of the subcluster slip along 
the contact surface between the subcluster plasma and the ambient plasma, 
the amplification of their strength is smaller than that in the 2D case. 
This weaker magnetic field, however, is sufficient to suppress the heat conduction 
across the front.

Since the area of the contact surface is larger in the 3D case than that in the 2D case.
Thus cold subcluster plasma can be heated more efficiently than in the 2D case.
We confirmed this by comparing results for model HC2 and HC3 (Figure \ref{fig:f6}).
Compared with model HC2 (2D, non-magnetic fields), the temperature 
profile for model HC3 (3D, non-magnetic fields) shows that the cold subcluster is 
heated up earlier. 
When magnetic fields exist, however, the difference becomes less prominent.

\ack

We thank T. Yokoyama for developments of the coordinated astronomical
numerical software (CANS) which include 2D and 3D MHD codes including heat conduction.
The development of CANS was supported by ACT-JST of Japan Science and Technology Corporation.
This work is supported by the priority research project in 
graduate school of Science and Technology, Chiba University (P.I., R. Matsumoto).
Numerical computations were carried out on VPP5000 
at the Astronomical Data Analysis Center, ADAC, of the National Astronomical 
Observatory of Japan.

\end{document}